\documentclass[letter,traditabstract]{aa}  

\usepackage{graphicx}
\usepackage{color}
\usepackage{txfonts}
\usepackage{natbib}
\definecolor{linkcolor}{rgb}{0.2, 0.55, 0.9}
\usepackage[breaklinks, colorlinks, citecolor=linkcolor]{hyperref}
\definecolor{rvdb}{rgb}{1.0,0.6,0}

\definecolor{jf}{rgb}{0,0.3,1}

\begin{document}

\title{Low Surface Brightness Galaxies in $z>1$ Galaxy Clusters:\\ HST approaches the Progenitors of Local Ultra Diffuse Galaxies}

    \titlerunning{Low Surface Brightness Galaxies in distant Galaxy Clusters probed with HST}

    \authorrunning{Aisha Bachmann et al.}
   \author{Aisha Bachmann
          \inst{1,2\thanks{As part of the 1st ESO Summer Research Programme}}, Remco F. J. van der Burg\inst{2}, J\'{e}r\'{e}my Fensch\inst{3,2},
          Gabriel Brammer\inst{4,5}, Adam Muzzin \inst{6}}

   \institute{Ruhr University Bochum, Faculty of Physics and Astronomy, Astronomical Institute, Universit\"atsstr. 150, 44801 Bochum, Germany\\
              \email{abach@astro.rub.de}
         \and
             European Southern Observatory, Karl-Schwarzschild-Str. 2, 85748, Garching, Germany \\
             \email{rvanderb@eso.org}
         \and           
            Univ. Lyon, ENS de Lyon, Univ. Lyon 1, CNRS, Centre de Recherche Astrophysique de Lyon, UMR5574, 69007, Lyon, France \\
            \email{jeremy.fensch@ens-lyon.fr}
            \and Cosmic Dawn Center (DAWN), Denmark
            \and Niels Bohr Institute, University of Copenhagen, Lyngbyvej 2, DK-2100 Copenhagen, Denmark
            \and Department of Physics and Astronomy, York University, 4700, Keele Street, Toronto, Ontario, ON MJ3 1P3, Canada
             }

     \date{Submitted 9 December 2020; accepted 15 January 2021 \vspace{0.1cm}} 


  \abstract{Ultra Diffuse Galaxies (UDGs), a type of large Low Surface Brightness (LSB) galaxies with particularly large effective radii ($r_\mathrm{eff}>1.5$ kpc), are now routinely studied in the local ($z$<0.1) universe. While they are found to be abundant in clusters, groups, and in the field, their formation mechanisms remain elusive and an active topic of debate. New insights may be found by studying their counterparts at higher redshifts ($z$>1.0), even though cosmological surface brightness dimming makes them particularly difficult to detect and study there. This work uses the deepest Hubble Space Telescope (HST) imaging stacks of $z$ > 1 clusters, namely: SPT-CL J2106-5844 and MOO J1014+0038. These two clusters, at $z$=1.13 and $z$=1.23, were monitored as part of the HST See-Change program. Compared to the Hubble Extreme Deep Field (XDF) as reference field, we find statistical over-densities of large LSB galaxies in both clusters. Based on stellar population modelling and assuming no size evolution, we find that the faintest sources we can detect are about as bright as expected for the progenitors of the brightest local UDGs.
   We find that the LSBs we detect in SPT-CL J2106-5844 and MOO J1014-5844 already have old stellar populations that place them on the red sequence. Correcting for incompleteness, and based on an extrapolation of local scaling relations, we estimate that distant UDGs are relatively under-abundant compared to local UDGs by a factor $\sim$3. A plausible explanation for the implied increase with time would be a significant size growth of these galaxies in the last $\sim$ 8 Gyr, as also suggested by hydrodynamical simulations.}


   \maketitle
%

\section{Introduction}
Dwarf galaxies are showing a vast range of properties in size and luminosity.  
Twenty particularly large (\textasciitilde 10 kpc) dwarfs with low surface brightness (LSB) were discovered through extensive photometric studies of the Virgo cluster by \citet{1984AJ.....89..919S}. 
An additional 27 examples were found in the Virgo cluster \citep{1988ApJ...330..634I} as well as in the Fornax cluster 
\citep{1988AJ.....96.1520F}. While these objects were found in high density environments, similar low surface brightness objects were also 
discovered in lower density environments such as the field \citep{1997AJ....114..635D,2019MNRAS.486..823R}.

More recently, after discovering 47 similar LSB objects (r\textsubscript{eff} \textasciitilde $3 - 10 ''$ , or r\textsubscript{eff} > 1.5 kpc, and $\mu (g,0) = 24 - 26$ mag arcsec$^{-2}$) in the Coma cluster, the term "Ultra Diffuse Galaxies" (UDGs) was introduced for these objects \citep{van_Dokkum_2015a}. Due to their projected density, and their spatial coincidence with the Coma cluster, 
\citet{van_Dokkum_2015a} concluded these objects to be a part of the Coma cluster, a statement that was confirmed by a follow-up spectroscopic study \citep{Dokkum_2015}.

Even though it is still a challenge to obtain redshift measurements of sizable samples of UDGs, their overdensity in galaxy clusters allows us to study their properties in such over-dense environments.
An approximately linear dependence between the abundance of UDGs in galaxy clusters, and the cluster mass was found \citep{van_der_Burg_2017,janssens17,romantrujillo2017}. 
UDGs within clusters appear to be found on the red sequence \citep{van_Dokkum_2015a,van_der_Burg_2016}, while UDG-like galaxies found in the field appear to be typically bluer \citep{2017ApJ...842..133L, prole19}.

Important open questions surrounding the study of UDGs are related to their origin, for which 
different theories have been proposed in the literature. 
\citet{van_Dokkum_2015a} suggest that (some) UDGs may have formed in haloes with masses similar to the Milky Way, but ``failed'' to form an L\textsubscript{*} galaxy. UDGs may also be the extremes in a continuous distribution in dwarf galaxy properties, having acquired their expanded sizes due to internal \citep{Amorisco_2016,Di_Cintio_2017}, or external \citep{Bennet_2018} processes.

Given the suggested low dark-matter content of some field UDGs \citep{van_Dokkum_2018}, they may also have formed as tidal dwarf galaxies \citep{Bennet_2018,Fensch_2019}.
Since these formation processes happen over different time scales, observing the evolving properties of UDGs may help distinguish between different scenarios.
To this end, we search for LSB galaxies in the two galaxy clusters SPTCL-2106-5844 ($z$ = 1.13) and MOO-1014+0038 ($z$ = 1.23). Using deep HST image stacks for those clusters, we reach the spatial resolution required to measure their sizes. We discuss our results in the context of the UDGs found in the local Universe.

This paper is organized as follows. Section \ref{data} provides an overview of the data we used. Section \ref{analysis} describes how we select our sample of UDGs. In Sect. \ref{results} we discuss our results regarding abundance and colour of the sample, and we summarize in Sect. \ref{conclusion}. 
We adopt the $\Lambda$CDM cosmology with $\Omega_m$ = 0.3, $\Omega_{\Lambda}$ = 0.7 and H\textsubscript{0} = 70 km s\textsuperscript{-1} Mpc\textsuperscript{-1}. At the redshift of our clusters, 1 arcsec corresponds to \textasciitilde 8.2 - 8.4 kpc. For stellar masses we assume the Initial Mass Function (IMF) from \citet{2003PASP..115..763C}. All magnitudes we quote are in the AB magnitude system.

\section{Data\label{data}}
We are making use of HST photometry taken as part of the See Change program (HST GO 13677, 14327; PI: Perlmutter), which targeted galaxy clusters in the range $1.13<z<1.75$. The main goal of their program was to find high-$z$ supernovae (SN) type \textbf{Ia}, and to use these to constrain the expansion rate of the universe. The survey strategy has therefore been to take data with a $\sim$ monthly cadence \citep[e.g.][]{williams20}. Rather than using the individual exposures, we use the image stacks, which reach a combined exposure time from 3.99 to 5.13 hours in F140W per cluster we examined. In this work we focus on the two lowest-$z$ clusters from their sample.

The first cluster we analyze is SPT-CL J2106-5844 (hereafter SPTCL-2106) at redshift $z$=1.13. It was discovered with the South Pole Telescope (SPT), thanks to its strong Sunyaev-Zel'dovich effect (SZ) signal, yielding a mass estimate of M\textsubscript{200} = (1.27 $\pm$ 0.21) $\times$ $10^{15}$ M$_\odot$ \citep{Foley_2011}.
The second cluster is MOO J1014+0038 (MOO-1014) at redshift z = 1.23, discovered by the Massive and Distant Clusters of WISE Survey (MaDCoWS) based on a rich overdensity of galaxies \citep{Gonzalez_2019}. It has a mass estimate of M\textsubscript{200} = (5.6 $\pm$ 0.6) $\times$  $10^{14}$ M\textsubscript{sun} and a strong SZ signature \citep{Brodwin_2015}.
RGB images of the two clusters are shown in Figs.~\ref{cluster1_rgb}~\&~\ref{cluster2_rgb}. 

   \begin{figure*}[h]
   \centering
   \includegraphics[width=19cm]{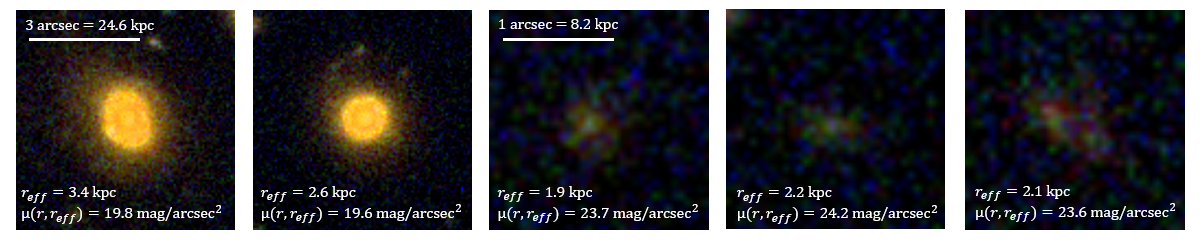}  
    \includegraphics[width=12cm]{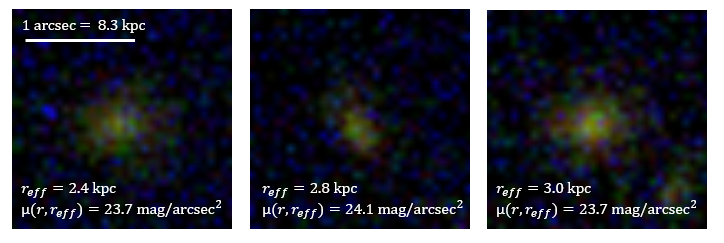}
   \caption{RGB (red = F140W, green = F105W, blue = F814W) images of members of the studied clusters. \textit{Top:} Two spectroscopically-confirmed bright members of SPTCL-2106 (on the left) and three LSB galaxies that are likely members of SPTCL-2106 (based on a reference field comparison, on the right). \textit{Bottom:} Three LSB galaxies that are likely members of the cluster MOO-1014 (based on a reference field comparison).}
              \label{sample_rgb}
    \end{figure*}

To create the full-depth mosaics of both clusters we start by aligning each of the multiple SN monitoring ``visits'' first internally to a catalog of sources detected in a single F140W visit and then globally to sources matched in the GAIA DR2 catalog \citep{gaiadr2}.  We use the \textsc{AstroDrizzle} software package \citep{2012drzp.book.....G} to identify and mask cosmic rays and bad pixels in the aligned individual exposures and perform source detection on the final combined F140W (WFC3/IR) mosaic generated with 60~mas pixels.  We further use F814W (WFC3/UVIS) stacks to provide basic colour information (or limits on the inferred colour based on the F814W detection limit). The combination of these filters bridge the 4000{\AA} break at the redshifts of our clusters, hence providing clues on stellar populations and a way to help assess the sample purity.

\subsection{Reference field}\label{sec:referencefield}
To estimate the level of contamination of the sample by foreground and background objects, we require a field survey with the same filter bands and image depth. We therefore utilise the 

data stacks taken in the Hubble eXtreme Deep Field\footnote{https://archive.stsci.edu/prepds/xdf/} (XDF). These deep stacks are composed of data from 19 different HST programs covering the Hubble Ultra Deep Field from 2002 to 2012. For details on the data reduction we refer to \citet{illingworth13}.

To ensure similar source detection limits as for the clusters, we add artificial noise
so that the recovered fraction of simulated sources were similar between the reference and cluster fields (see Sect. \ref{sims}).

    \begin{figure*}[h]
   \centering
   \includegraphics[width=18.0cm]{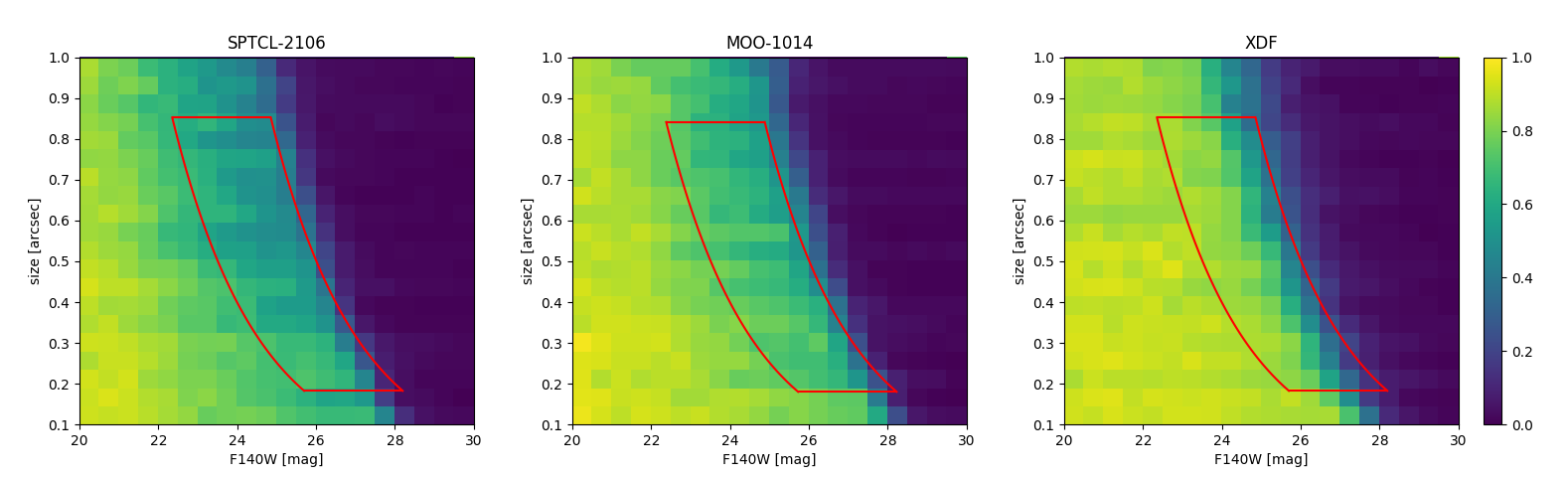}
   \caption{\textit{Left:} Recovery fractions for the simulated sources in SPTCL-2106. \textit{Middle:} Same for MOO-1014. \textit{Right:} Same for the XDF reference field, with noise level matched to the cluster fields. 
   The colourbar shows the recovery fractions. 
   While the detection limits are comparable between the panels, the XDF shows overall a better recovery fraction for brighter targets than in the clusters due to a higher source crowding in the cluster fields. Also highlighted is the regime of objects with a surface brightness from 24.0 to 26.5 mag arcsec$^{-2}$  in F140W and a radius from 1.5 to 7.0 kpc.}
              \label{Histo}%
    \end{figure*}
%

\section{Analysis\label{analysis}}
\subsection{Source detection}
Sources are detected by running \texttt{SExtractor} \citep{1996A&AS..117..393B} on the F140W images of the clusters and the XDF.
The parameters used to ensure optimal detection of faint and extended sources in the clusters can be found in Appendix \ref{parameters}, and an identical setup was used for the XDF images.
We only consider sources in the relatively central regions of the cluster stacks, where exposure time is nearly uniform (cf.~Figs.~\ref{cluster1_rgb}~\&~\ref{cluster2_rgb}).
Examples of detected objects in each of the two clusters are shown  Fig.~\ref{sample_rgb}.

\subsection{Structural parameters}
For the detected sources, structural parameters such as magnitude, radius, ellipticity and S\'ersic index of the detected objects are determined using GALFIT \citep{2002AJ....124..266P} on the F140W image after masking neighbouring objects that are detected by \texttt{SExtractor}. GALFIT is also allowed to simultaneously fit a constant value to the sky background to improve the overall fit. 
In order to measure reliable colours, the F814W fluxes of the detected objects are measured on the corresponding stack by forcing GALFIT to use the morphological parameters obtained from the F140W stack, and only fit the flux normalisation. 

To estimate the measurement uncertainty, the GALFIT models are injected on a hundred different random locations in the corresponding cluster, requiring that there were no prior detections, and measured exactly as before. In this way the 1-$\sigma$-uncertainties for size and magnitude in F140W, and magnitude in F814W, are obtained. We note that GALFIT measured a slightly lower flux (by about 0.2 mag) and smaller size (by about 0.3 kpc) than the simulated inputs. In the following, we correct for this small bias.

\subsection{Image simulations\label{sims}}
To assess the completeness of our UDG progenitor selection, we perform all processing steps also on a range of image simulations. For this we inject objects with S\'ersic profiles on random locations in the HST stacks. We choose a constant S\'ersic-$n$ parameter of unity, which corresponds to typical light profiles of UDGs measured in the local Universe \citep[e.g.][]{van_Dokkum_2015a,koda15,van_der_Burg_2016}. Sizes are drawn uniformly between 0\farcs1 and 1\farcs0, and ellipticities $f$, defined as $f = 1 - b/a$  with b/a the axis ratio, uniformly between 0.0 and 0.2. Each step was performed identically on the cluster- and on the reference field image. 

Fig. \ref{Histo} shows the recovered fractions of the inserted objects, for the two clusters and the reference field. 
While the detection limits in the different panels looks similar overall, we note that there is a substantial difference between the clusters and the reference field. Even for relatively bright sources, the recovery fraction is lower in the cluster than in the field. This is expected given the relatively high number of large and bright sources crowding the cluster stacks. 
These recovery fractions are used to determine the limits of our analysis, and to correct the detected sources for incompleteness, both due to limiting depth and due to crowding/obscuration.

\subsection{Sample selection \label{selection}}
While the definition of UDGs is rather arbitrary, we initially filter the sample by using a definition similar to that used in the Local Universe:
\begin{itemize}
    \item surface brightness in F140W $\geq$ 24.0 mag arcsec$^{-2}$ (i.e., not corrected for surface brightness dimming),
    \item effective radius between 1.5 and 7.0 kpc,
    \item S\'ersic index $\leq$ 4.0 to increase the sample purity in favour of sources reminiscent of UDGs in the local universe,
    \item distance between \texttt{SExtractor} detection and GALFIT position of measurement < 3.0 pixels to ensure that both consider the same object.
\end{itemize}

A fainter limit for the surface brightness is not included as one purpose of the study is to test the detection limits of LSB galaxies with the available data.
After detecting sources in direction of the cluster we assume that each source is at the redshift of the cluster to infer their physical parameters. This is a valid assumption since we subsequently perform a statistical subtraction of fore- and background objects, thereby making the same assumption for sources detected in the reference field. 

After discarding failed detections with a by-eye scan we find 95 objects (10 discarded) in SPTCL-2106 (70 in reference field, 6 discarded) and 111 objects (10 discarded) in MOO-1014 (72 in reference field, 9 discarded). We note that the number of objects in the reference field depends on the assumed angular diameter distance, and thus the redshift of the cluster.
We conservatively only consider sources that would have had a detection probability of at least 50$\%$ in the reference field for both cluster and reference field objects and only count the objects above this limit to ensure a reasonable completeness correction. For the remaining sources we apply a completeness correction that is thus at most a factor 2 based on the determined recovery fractions (see Sect. \ref{sims}). We also apply a correction based on the different sky area covered by the cluster and reference field. We are left with a statistical count of 99 $\pm$ 10 objects in SPTCL-2106 (67 $\pm$ 8 in reference field) and 90 $\pm$ 10 in MOO-1014 (70 $\pm$ 8 in reference field).

Colour-Magnitude diagrams for all the detected sources falling within the 50$\%$-limit (see Fig.~\ref{color_magall}) in both clusters and the reference field are shown in the Appendix \ref{app2}.
We subtract the reference field galaxies from the nearest cluster object in regards of colour (F814W-F140W) and magnitude (F140W). For this we use the colour and magnitude measurements for each cluster object as coordinates and subtracted the reference field object from the nearest cluster object. The relative weights, which include corrections for incompleteness and different covered sky areas, are taken into account. For more details on the subtraction of the reference field we refer to \citet{van_der_Burg_2016}.
After the subtraction we are left with a statistical count of 32 $\pm$ 13 in SPTCL-2106 and 20 $\pm$ 13 in MOO-1014.

  \begin{figure*}
   \centering
    \includegraphics[width=18cm]{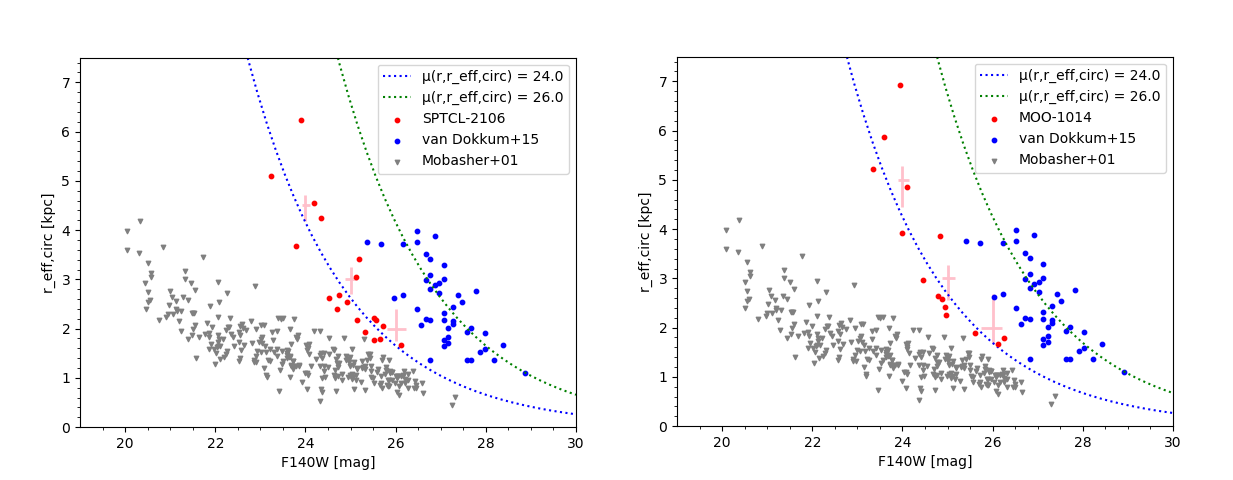}
      \caption{A selection by size and apparent magnitude, in F140W, of our LSB galaxies. The panels show the different clusters. \textit{Red:} LSB galaxies detected following our selection criteria and with a weight higher than 0.5. \textit{Blue and Grey:} the samples by \citet{van_Dokkum_2015a} and \citet{Mobasher_2001}, both shifted to our observed redshift by accounting for an E+K correction (see Sect. \ref{results}). We plot curves of constant surface brightness, $\mathrm{\mu(r_{eff,circ}})$ = 24.0, 26.0 mag arcsec$^{-2}$ evolved from the Coma cluster redshift to the high-$z$ clusters. Average errorbars for our sample are shown.}
               \label{rad_mag}
   \end{figure*}
   
  \begin{figure}
   \centering
   \includegraphics[width=\hsize]{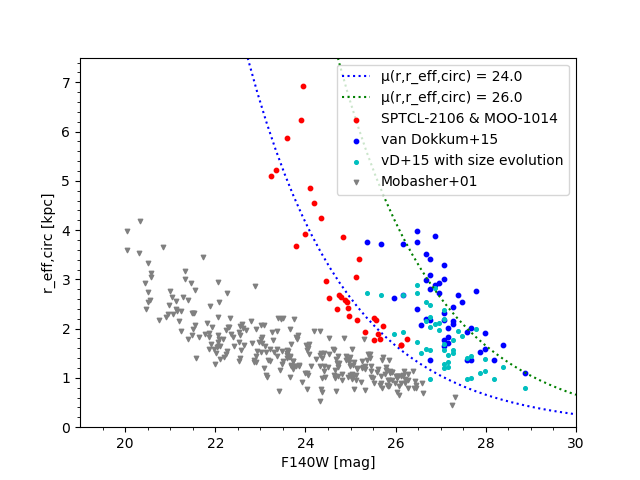}
      \caption{Same es Fig. \ref{rad_mag} but for both clusters combined and with the sample by \citet{van_Dokkum_2015a} with the size evolution taken out. \textit{Cyan:} the sample by \citet{van_Dokkum_2015a}, where the suggested size evolution since $z\sim 1$ (cf.~Sect. \ref{abundance}) is taken out.
              }
         \label{evolved}
   \end{figure}  
 \begin{figure*}
   \centering
    \includegraphics[width=18cm]{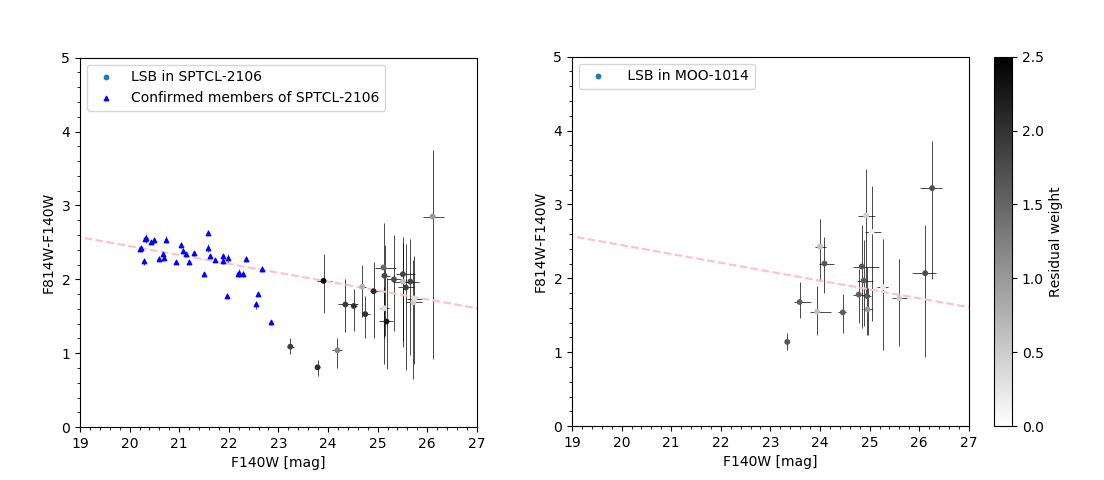}
      \caption{Colour-Magnitude diagrams for the sample of LSB candidates in the clusters after correcting for fore- and background interlopers and incompleteness. The colourbar gives the residual weight of the data points (as detailed in Sect. \ref{selection}). The left panel shows the cluster SPTCL-2106 and the right panel the cluster MOO-1014. Additionally shown in the left panel are the positions in the colour-magnitude diagram of spectroscopically-confirmed members of SPTCL-2106, and an extrapolation of the red sequence as defined by the confirmed members of SPTCL-2106, ignoring some of the data points that may be part of a bluer/starforming cluster population to guide the reader's eye. This line has a slope of -0.12. 
              }
         \label{color_mag}
   \end{figure*}
   
\section{Results and Discussion\label{results}}

\subsection{Comparison with local UDGs \label{Compare}}
To be able to compare our LSB candidates to likely progenitors of UDGs studied in the local universe, we evolve local UDGs back to the redshift of our clusters following a simple stellar population model. The model assumes a passively-evolving stellar population that was formed at $z_{\mathrm{form}}=1.5$, which is in line with intermediately old ages of UDGs measured locally \citep[][]{ferremateu18,ruizlara18,Fensch_2019}. It is based on stellar population synthesis models from  \citet{bc03}, a star formation history $\mathrm{SFR} \propto e^{-t/\tau} $ with a short e-folding time of $\tau$ = 10 Myr, a \citet{2003PASP..115..763C} initial mass function and no dust extinction. We use the magnitudes, physical sizes and filters of \citet{van_Dokkum_2015a} as anchor point, and estimate how those UDGs would appear at the redshift of our cluster as observed through the WFC3/F140W filter. This estimate accounts, by construction, for surface brightness dimming. The 47 UDG candidates are shown in Fig. \ref{rad_mag}.
Additionally we evolved the dwarf and giant galaxies found by \citet{Mobasher_2001} in the Coma cluster back to the redshift of our clusters in the same way and plotted them as well in Fig. \ref{rad_mag}.

Figure \ref{rad_mag} shows that the LSB samples of our clusters lie in the area between the samples by \citet{van_Dokkum_2015a} and \citet{Mobasher_2001}, making them fainter than the progenitors of normal dwarf and giant galaxies in the Coma cluster, and almost as faint as the expected progenitors of the UDGs studied by \citet{van_Dokkum_2015a}. We can see a small overlap between our samples and the compared objects from both other studies.
As a reference, we also plot the curves of constant surface brightness, $\mu(r,r_\mathrm{eff,circ}) = 24.0,26.0$ mag arcsec$^{-2}$, which is a common selection boundary for UDGs in the local universe, also evolved to the redshifts of our clusters. This indicates that only half of the objects we are able to detect in both clusters could classify as progenitors of the brightest UDGs known in the local universe, based on this evolution model.

We note that, while projecting the local Coma galaxies back to higher redshift, we have only evolved their fluxes and ignored any potential size evolution.However, numerical simulations suggest that a typical UDG may see its radius increase with age from around 2.5 - 3 kpc at $z$ = 1 to 4 - 5 kpc at $z$ = 0 \citep{Martin_2019,wright2020formation}. Accounting for such an expansion would bring the data points from \citet{van_Dokkum_2015a}, when evolved to the redshift of our clusters, downward, and thus closer to our data points.
A possible size evolution is described in Sect. \ref{abundance} and the sample by \citet{van_Dokkum_2015a} effected by it is plotted  in Fig. \ref{evolved}.

\subsection{Colour \label{colour}}
Figure \ref{color_mag} shows the colours and magnitudes of the sample of LSB galaxies. The residual weight shows the number of galaxies we expect in the observed cluster area per detected object after accounting for the subtraction of the reference field.
Also shown in Fig. \ref{color_mag} are the positions in the colour-magnitude diagram of galaxies which were spectroscopically confirmed as members of SPTCL-2106 \citep[by the GOGREEN collaboration,][]{balogh20}. The colours were measured in the same filter bands and following an identical method as for the LSB candidates. 
Ignoring some of the data points that may be part of a bluer/star-forming cluster population, we note that the bulk of the LSB galaxies lie on an extended red-sequence, shown as a pink-dashed line, as defined by the brighter cluster galaxies. We note that the slope of this estimate red-sequence, -0.12, is consistent with $z\sim 1$ estimates by e.g.~\citet{bell04}. No similar population of bright cluster members has been spectroscopically identified for MOO-1014, so that we plot the same red-sequence estimate for this cluster, ignoring the small redshift difference between the two clusters. The LSB galaxies of SPTCL-2106 and MOO-1014 show colours that are consistent with the red sequence, suggesting that these galaxies are likely quenched and have thus already stopped forming stars.

\subsection{Abundance comparison with local universe UDGs}
\label{abundance}
To put the measured abundance of LSB galaxies in our clusters into context, we compare it to the abundance of UDGs in local clusters, within projected $R<R_{200}$. For this, we have to make assumptions regarding the underlying magnitude and size distribution of dwarf galaxies, and how these evolve with redshift. We assume a flat magnitude distribution for different size bins \citep[consistent with what is observed in the Coma cluster, by][]{2019ApJ...875..155D}, and the same size distribution as measured for UDGs in local clusters \citep[for radii between 1.5 and 7.0 kpc][]{van_der_Burg_2016}.
Based on Fig.~\ref{Histo} we can also assume that our samples in the range with surface brightness from 24.0 to 26.5 mag in F140W and a radius from 1.5 to 7.0 kpc are mostly complete for the parameter range studied of both clusters.

The available \textit{HST} imaging does not allow us to probe radii out to $R_{200}$, but only to radii corresponding to $\sim 0.35\cdot R_{200}$ for SPTCL-2106 and $\sim 0.50\cdot R_{200}$ for MOO-1014. To correct for the missing area, we assume that LSBs approximately trace the overall matter distribution in the cluster, which is described by an NFW \citep{NFW} profile with concentration $c_{200}$=3 \citep[e.g.][]{duffy08}. Integrating this profile along the line-of-sight indicates that we probe a fraction of $\sim 0.4\pm 0.1$ of the LSB population in SPTCL-2106 and $\sim 0.55\pm 0.1$ in MOO-1014.
Based on the assumed magnitude and size distribution\footnote{We considered uncertainties in the assumed size distribution \citep[as measured in][]{van_der_Burg_2016} and magnitude distribution \citep[as measured in][]{2019ApJ...875..155D}, finding that these affect our estimated number of high-z UDGs by at most 14\%, hence not impacting our conclusions.}, and after applying the needed correction for missed area, we would estimate a total number of 80 $\pm$ 38 UDGs in SPTCL-2106 and 36 $\pm$ 25 UDGs in MOO-1014. Studies of local clusters suggest an abundance of $\sim$100-200 in clusters of this mass, being three times higher than our best estimate for the clusters we study. This thus implies a substantial increase in the UDG abundance with time since $z\sim 1$.
A possible explanation for this implied evolution is that we are assuming the UDG progenitors to be of the same size as local Universe UDGs. If their progenitors were actually smaller at $z$ = 1 \citep[as suggested by several simulations, cf.][]{Martin_2019,wright2020formation} they would not fall into our selection criteria and thus be missed in the current analysis.

Assuming the size distribution of UDGs in the local universe, as described by the power law shown in Fig. 7 of \citet{van_der_Burg_2016}, we find that a size growth by a factor $\sim$1.4 of all galaxies may have boosted the number of galaxies classified as UDGs by a factor $\sim$3 since $z\sim 1$. Hydrodynamical simulations by \citet{Martin_2019} would predict a slightly larger size growth by a factor $\sim$1.8. This suggests that size evolution is sufficient to explain the observed underabundance of UDGs.

\section{Summary and outlook}
\label{conclusion}
This paper studies the abundance of LSB galaxies in two $z > 1$ clusters, SPTCL-2106 and MOO-1014, down to the detection limit of the available deep HST imaging data. We 
correct for background objects by comparing the detections with those measured in the XDF as reference field.
Simulations are run to estimate completeness limits, and to tailor the depth of the reference field to the cluster imaging.
We summarise our main conclusions as follows:

   \begin{itemize}
   
      \item Within the parameter space we defined, we find a statistical overdensity of 32 $\pm$ 13 LSB galaxies in SPTCL-2106 and 20 $\pm$ 13 in MOO-1014.
      
      \item We find the colours of those LSB galaxies in SPTCL-2106 and MOO-1014 to be consistent with an extension of the red sequence, as defined by spectroscopically identified brighter cluster members. This suggests that the LSB galaxies in both clusters are already evolving passively. 
      
      \item Based on a simple stellar population evolution model, we compare our detected LSB galaxies with the expected progenitors of local UDGs in the Coma cluster. This suggests that the faintest sources we can detect approximate the expected progenitors of local UDGs. 
      
      \item Based on an extrapolation, motivated by local scaling relations, we estimate an overall abundance of 80 $\pm$ 38 UDGs in SPTCL-2106 and 36 $\pm$ 25 UDGs in MOO-1014. We note that this is about three times lower than the abundance of UDGs in local galaxy clusters having similar masses.
      
      \item One way to interpret the implied evolution is by assuming a substantial size growth of dwarf galaxies since $z\sim 1$, which would then increase the numbers of those that classify as UDGs. As we discuss, this is qualitatively consistent with results from hydrodynamical simulations.
      
   \end{itemize} 
  
We stress that this study uses the deepest data available for galaxy clusters at $z$ > 1 that still allow to spatial resolve galaxies of 1-2 kpc, and thus reaches the limit of current instrumentation. For further studies on the existence and properties of distant UDGs, data with higher spatial resolution and depth is needed, being within reach of the next generation of telescopes.

\begin{acknowledgements}
We thank the anonymous referee for their useful comments that substantially clarified the paper.
AB acknowledges a 6-week ESO summer studentship during which a substantial part of this research was done. 

\end{acknowledgements}

%
%

\bibliographystyle{aa} 
\bibliography{MasterRefs} 

\begin{appendix}
\section{RGB images of the clusters}

 \begin{figure*}[h]
   \centering
   \includegraphics[width=15cm]{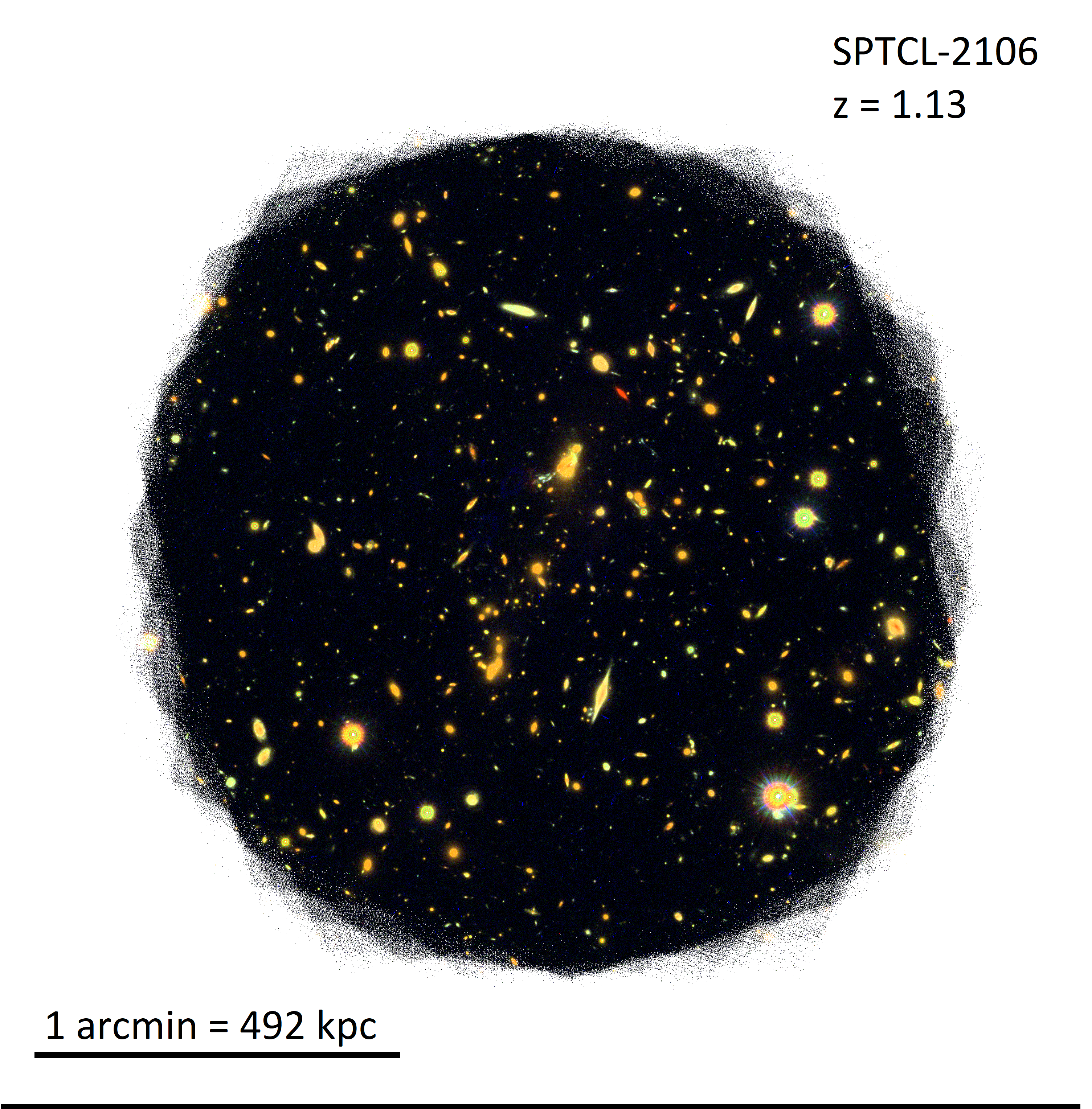}
   \caption{RGB (red = F140W, green = F105W, blue = F814W) image of the cluster SPTCL-2106.}
              \label{cluster1_rgb}%
    \end{figure*}
    
 \begin{figure*}[h]
   \centering
   \includegraphics[width=15cm]{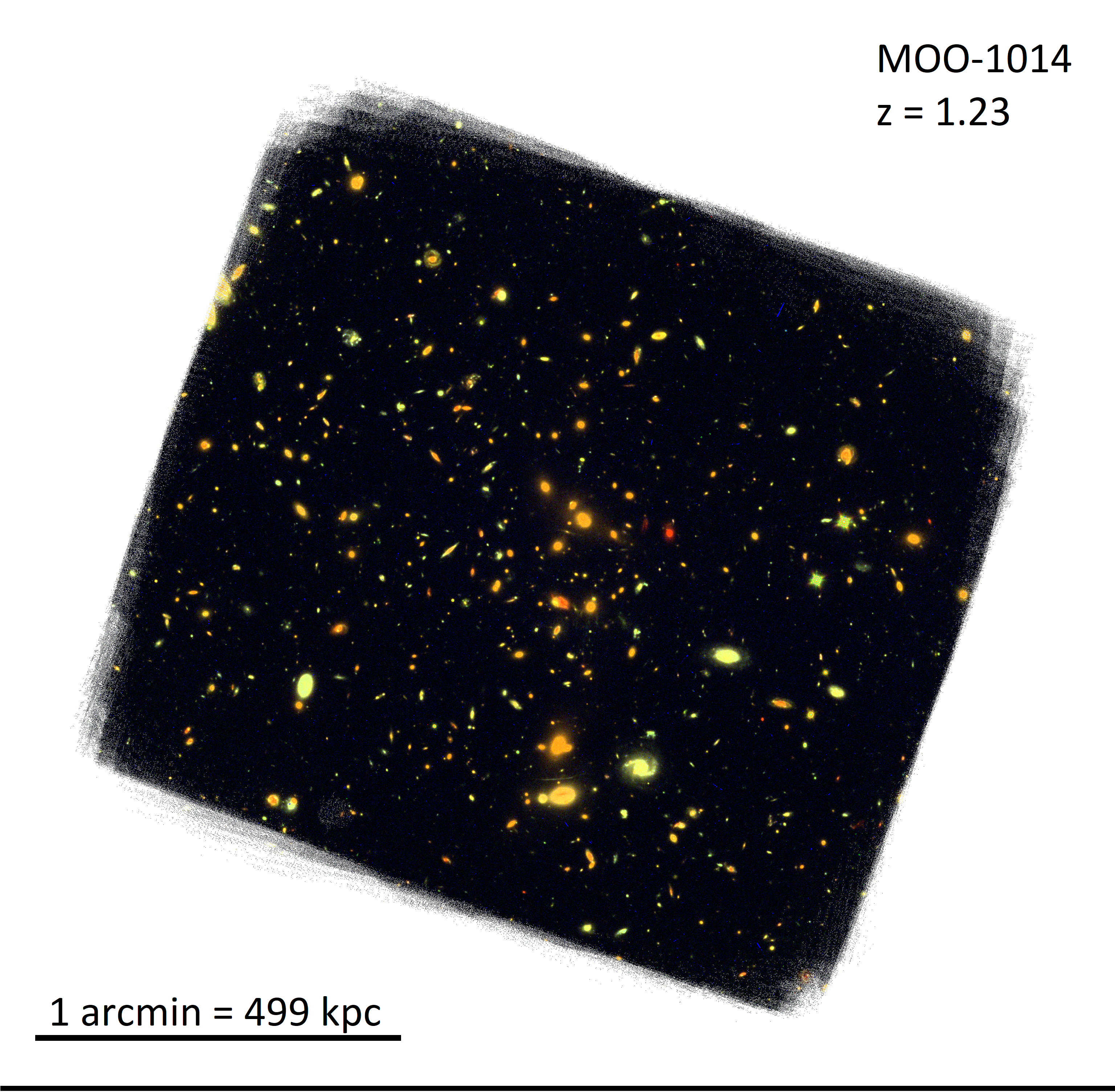}
   \caption{RGB (red = F140W, green = F105W, blue = F814W) image of the cluster MOO-1014.}
              \label{cluster2_rgb}%
    \end{figure*}

\section{\texttt{SExtractor} parameters}\label{parameters}

\begin{table}[ht]
\centering
\caption{\texttt{SExtractor} parameters used. All other parameters were left to their defaults.}
\begin{tabular}{l l}
\hline \hline
Parameter & Value \\
\hline
 DETECT\_MINAREA & 7 \\
 DETECT\_THRESH &  1.1 \\
 ANALYSIS\_THRESH &  1.1 \\
 BACK\_TYPE & MANUAL \\
 BACK\_VALUE & 0 \\
 FILTER\_TYPE & GAUSSIAN \\
 FILTER & default \\
 \hline
\end{tabular}
\end{table}

\section{Additional Figures}\label{app2}

\begin{figure*}[]
   \centering
    \includegraphics[width=18cm]{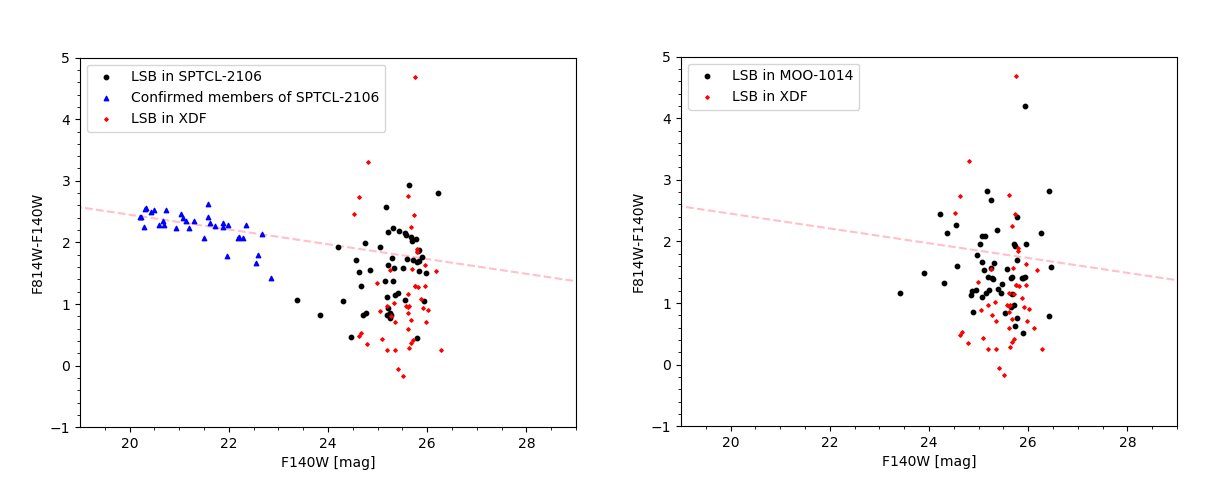}
      \caption{Colour-Magnitude diagrams for the sample of LSB candidates in the clusters and the reference field that fall within the 50\% detection limit. These are the raw numbers, without accounting for incompleteness, or background interlopers. The left panel shows the cluster SPTCL-2106 and the right panel shows the cluster MOO-1014. Additionally shown in the left panel are the positions in the colour-magnitude diagram of spectroscopically-confirmed members of SPTCL-2106 and the same extrapolation of the red-sequence as described in Fig. \ref{color_mag}.
              }
         \label{color_magall}
   \end{figure*}

\end{appendix}

\end{document}